\documentclass[aps,pre,preprint,superscriptaddress]{revtex4-1}
\usepackage{times}
\usepackage{xcolor}
\definecolor{boxcolor}{RGB}{255,255,255}
\begin{document}

\title{Mathematical foundations of moral preferences}

\author{Valerio Capraro}
\email{v.capraro@mdx.ac.uk}
\affiliation{Department of Economics, Middlesex University, The Burroughs, London NW4 4BT, U.K.}

\author{Matja{\v z} Perc}
\email{matjaz.perc@gmail.com}
\affiliation{Faculty of Natural Sciences and Mathematics, University of Maribor, Koro{\v s}ka cesta 160, 2000 Maribor, Slovenia}
\affiliation{Department of Medical Research, China Medical University Hospital, China Medical University, Taichung 404332, Taiwan}
\affiliation{Alma Mater Europaea ECM, Slovenska ulica 17, 2000 Maribor, Slovenia}
\affiliation{Complexity Science Hub Vienna, Josefst{\"a}dterstra{\ss}e 39, 1080 Vienna, Austria}

\begin{abstract}
One-shot anonymous unselfishness in economic games is commonly explained by social preferences, which assume that people care about the monetary payoffs of others. However, during the last ten years, research has shown that different types of unselfish behaviour, including cooperation, altruism, truth-telling, altruistic punishment, and trustworthiness are in fact better explained by preferences for following one's own personal norms -- internal standards about what is right or wrong in a given situation. Beyond better organising various forms of unselfish behaviour, this moral preference hypothesis has recently also been used to increase charitable donations, simply by means of interventions that make the morality of an action salient. Here we review experimental and theoretical work dedicated to this rapidly growing field of research, and in doing so we outline mathematical foundations for moral preferences that can be used in future models to better understand selfless human actions and to adjust policies accordingly. These foundations can also be used by artificial intelligence to better navigate the complex landscape of human morality.
\end{abstract}

\date{\today}

\maketitle

\section{Introduction}\label{intro}

Most people are not completely selfish. Given the right circumstances, they are happy to give up a part of their benefit to help other people or the society as a whole. Psychologists and economists have long observed that some people act unselfishly even in one-shot anonymous interactions, when there are no direct or indirect benefits for doing so \cite{rapoport1965prisoner,engel_ee11}. The question is why? Understanding what motivates people to act unselfishly in one-shot, anonymous interactions is of great theoretical and practical importance. Theoretically, it may lead to a more complete and precise mathematical framework to formalise human decision-making, while practically, it may suggest policies and interventions to promote unselfish behaviour, with the ultimate goal of improving our societies.

To study one-shot unselfishness, behavioural scientists usually turn to laboratory experiments using economic games, in which experimental subjects have to make monetary decisions that involve various forms of other-regarding behaviour. In this context, and throughout this review, selfishness and other-regarding behaviour is defined with respect to monetary payoffs. Clearly, a behaviour that is unselfish from the point of view of monetary outcomes may turn out to be selfish from a more general perspective that takes into account also psychological benefits and costs. For example, some people may engage in unselfish behaviour to decrease negative mood \cite{cialdini1973transgression} or increase positive feelings \cite{andreoni1990impure}. Therefore, in the last decades, behavioural scientists have been trying to mathematically explain unselfish behaviour by means of a utility function that depends on factors other than solely the monetary payoff of the decision maker. Based on empirical data scholars have initially advanced the explanation that human unselfishness in one-shot anonymous interactions is primarily driven by people not caring only about their own monetary payoff, but caring, at least to some extent, also about the monetary payoffs of the other people involved in the interaction \cite{ledyard1994public, levine1998modeling, fehr1999theory, bolton2000erc, andreoni2002giving, charness2002understanding}.

However, about fifteen years ago, this \emph{social preference hypothesis} came under critique because some experiments showed that two particular forms of unselfish behaviour, altruistic punishment and altruism (see Table 1 for these definitions), could not be entirely explained by preferences defined solely over monetary outcomes. In 2010, building on work on the effect of social norms on people's behaviour \cite{smith2010theory,durkheim2017regles, parsons1937structure, geertz1973interpretation, schwartz1977normative, elster1989social, cialdini1990focus, bicchieri2005grammar,bicchieri2016norms,hawkins2019emergence}, Bicchieri and Chavez \cite{bicchieri2010behaving} proposed to explain altruistic punishment assuming that people have preferences for following their ``personal norms'' (what they personally believe to be the right thing to do) beyond the monetary consequences that this action brings about. Subsequently, Krupka and Weber \cite{krupka2013identifying} proposed to explain altruism using ``injunctive norms'' (what one believes others would approve/disapprove); however, in their analysis, they did not consider a potential role of personal norms. In the last five years, numerous other experiments challenged social preference models in several behavioural domains, other than altruistic punishment and altruism \cite{biziou2015does,schram2015inducing,kimbrough2016norms, eriksson2017costly, capraro2018right, tappin2018doing,capraro2019power,huang2019choosing}; moreover, the best interpretation of these results turns out to be in terms of personal norms, rather than other types of norms. Namely, the best way to organise these results is through the moral preference hypothesis, according to which people have preferences for following their personal norms, beyond the economic consequences that these actions bring about. This framework organises several forms of one-shot, anonymous unselfish behaviour, including cooperation, altruism, altruistic punishment, trustworthiness, honesty, and the equality-efficiency trade-off. We note at this stage that personal norms are not universally given. They certainly depend on the culture; for example, they can come from the internalisation of cultural values \cite{schwartz1977normative}. But they can also depend on the individual; anecdotal evidence suggests that, even within the same family, there might be people with different beliefs about what is right or wrong in a given situations. We will discuss this in more details in Section \ref{suse: phenotype}.

The moral preference hypothesis also holds promise of being very useful in practice. The idea is simple. If people care about doing the right thing, then just providing cues that make the rightness of an action salient should work just fine in promoting desirable behaviour. In fact, research has already demonstrated the applicability of this approach outside of the laboratory, showing in particular that nudges towards doing the right thing can increase charitable donations \cite{capraro2019increasing}.

In the light of ample empirical research supporting the moral preference hypothesis, theoretical research aiming to formalise human decision-making by means of a mathematical framework is also at a crossroads. On the one hand, the traditional approach involving monetary payoffs has worked well in explaining many challenging aspects of pro-social behaviour. But on the other, experiments indicate that there are likely hard boundaries to this simplistic approach, which will thus have to be amended by more avant-garde concepts, including formalising the intangibles of moral psychology and philosophy.

Here we review this rapidly growing field of research within the following sections. Section \ref{unselfishness} reviews the main economic games that have been developed to study one-shot unselfishness. Section \ref{social preferences} reviews social preference models, as earlier attempts to explain unselfishness in one-shot economic games within a unified theoretical framework. This section also describes a number of experiments that violate social preference models. Section \ref{norm} shows how these experiments can be organised by general moral preferences for doing what one believes to be the right thing. Section \ref{applications} focuses on practical applications of the moral preference hypothesis. Section \ref{moral preferences} reviews the models of moral preferences that have been introduced so far and proposes a new model that explicitly takes into account the importance of personal norms. Lastly, Section \ref{future work} outlines a number of key questions for future work, while Section \ref{conclusion} summarises the main conclusions.

Taken together, this review thus outlines a mathematical formalism for morality, which shall inform future models aimed at better understanding selfless actions as well as artificial intelligence that strives to emulate counterintuitive human decision-making.

\section{Measures of unselfish behaviour}\label{unselfishness}

There are various forms of unselfish behaviour. For example, giving money to a homeless person on the street is, in principle, quite different from collaborating with a colleague on a common project, or from telling the truth when one is tempted to lie. To take this source of heterogeneity into account, scholars have developed a series of simple games and decision problems that are meant to prototypically represent different types of unselfish behaviour. These are simple scenarios in which experimental subjects can make decisions that have real consequences. To incentivise these decisions, behavioural scientists usually use monetary payoffs (at least among adult subjects, whereas other forms of remuneration, such as stickers, might be more effective among children).

In this review, we will be mainly focused on one-shot decisions that are \emph{purely} unselfish, meaning that they bring no monetary benefit to the decision maker (and possibly bring a cost), no matter the beliefs of the decision maker regarding the behaviour of other people involved in the interaction. Specifically, we measure altruistic behaviour using the dictator game (see Table 1 for all the definitions), cooperative behaviour in pairwise interactions using the prisoner's dilemma, truth-telling using the sender-receiver game,  the tradeoff between equality and efficiency using the trade-off game, trustworthiness using player 2 in the trust game, and altruistic punishment using player 2 in the ultimatum game. In the last section we will also briefly consider decisions that are \emph{strategically} unselfish, such as trust (player 1 in the trust game) and strategic fairness (player 1 in the ultimatum game), which might actually maximise the payoff of the decision maker, depending on their beliefs about the behaviour of the second player. The distinction between pure unselfishness and strategic unselfishness generalises the distinction between pure cooperation and strategic cooperation, introduced by Rand in his meta-analysis \cite{rand2016cooperation}, where “pure cooperation” was defined as paying a cost to benefit another person, regardless of the behaviour of the other person, as opposite of “strategic cooperation”, which might maximise the cooperator’s payoff, depending on the other person’s behaviour.

\section{Social preferences and their limitations}\label{social preferences}

Behavioural scientists have long recognised that some people do act unselfishly even in one-shot anonymous interactions. For example, the first comprehensive empirical work on the one-shot prisoner's dilemma dates back to 1965 \cite{rapoport1965prisoner}. Formal frameworks to explain one-shot unselfishness have a more recent history, starting in 1994, when Ledyard observed that cooperation, altruism, and altruistic punishment could be explained by assuming that people maximise a utility function that depends not only on their own monetary payoff, but also on the total monetary payoff of the other people that are involved in the interaction \cite{ledyard1994public}. See Table 2 for the exact mathematical definition. Since then, several models have been introduced. In 1998, Levine \cite{levine1998modeling} proposed a utility function in which the level of altruism depends on the level of altruism of the other players. Subsequently, in 1999, Fehr and Schmidt \cite{fehr1999theory} proposed a framework according to which players care about minimising inequities. In 2000, Bolton and Ockenfels \cite{bolton2000erc} followed a similar idea and introduced a general inequity aversion model, in which the utility of an action depends negatively on the distance between the amount of money the decision maker gets if that action is implemented and the amount of money the decision maker would get if the equal allocation were implemented. The authors proposed an explicit mathematical formula only for the case of $n=2$ players. In 2002, Andreoni and Miller \cite{andreoni2002giving} estimated the behaviour of experimental subjects in a number of dictator game choices using a specific utility function taking into account altruistic tendencies as well as potential convexity in the preferences. In the same year, Charness and Rabin \cite{charness2002understanding} introduced a general utility function which, depending on the relative relationship between its two parameters, can cover several cases, including competitive preferences, inequity aversion preferences, and social efficiency preferences. We refer to Table 2 for the exact functional forms. (Besides these models, scholars have also studied models that can be applied to specific subsets of one-shot anonymous interactions (e.g., \cite{andreoni1990impure}). In this review, we focus on models that can be applied to any one-shot anonymous interaction involving unselfish behaviour).

While differing in many details, all social preference models share one common property: they assume that the utility of a decision maker is a function of the monetary payoffs of the available actions. This assumption came under considerable criticism for the first time in 2003 when Falk, Fehr and Fischbacher \cite{falk2003nature} showed that rejection rates in the ultimatum game depend on the choice set available to the proposer. Specifically, the split (8,2) --- 8 to the proposer and 2 to the responder --- is more likely to be accepted in ultimatum games in which the only other choice available to the proposer is (10,0), compared to ultimatum games in which the only other choice available to the proposer is (5,5). Therefore, responders prefer accepting (8,2) over rejecting it in the former case, but they prefer rejecting (8,2) over accepting it in the latter one, despite the fact that these choices have the same monetary consequences in the two cases. Clearly, this cannot be explained by any model of social preferences. See \cite{bicchieri2010behaving,eriksson2017costly} for conceptual replications.

Shortly after, in 2005, Uri Gneezy introduced the sender-receiver game \cite{gneezy2005deception}. In his experiments, decision makers were less likely to implement an allocation of money when implementing this allocation also required misreporting a private information. Also this finding cannot be explained by any model of social preferences and, more generally, also not by any utility function that depends only on the monetary payoffs that are associated with the available actions. This thus indicates that (some) people have an intrinsic cost of lying, which goes beyond their preferences toward monetary outcomes. To further support this interpretation, several scholars have independently studied the sender-receiver game in contexts in which lying would benefit both the sender and the receiver to the same extent. This case is particularly important because, when the benefit for the sender is equal to the benefit for the receiver, all social preference models predict that the totality of people would lie. However, this prediction turned out to be violated in experiments, which showed that a significant proportion of people tell the truth \cite{cappelen2013we, erat2012white, biziou2015does}.

Subsequently social preference models came under critique also in one of the behavioural domains in which they had been most successful, namely in research involving the dictator game. Dana, Cain and Dawes \cite{dana2006you} and Lazear, Malmendier and Weber \cite{lazear2012sorting} observed that some dictator game givers would prefer to altogether avoid the dictator game interaction if given the chance. These people thus preferred giving over keeping in a context in which they were forced to play the dictator game, but preferred keeping over giving in a context in which they could choose whether to play the dictator game or not. This finding, as in the preceding examples, cannot be explained by any utility function that is based solely on monetary outcomes.

For the same game, and along similar lines, List \cite{list2007interpretation}, Bardsley \cite{bardsley2008dictator}, and Cappelen et al. \cite{cappelen2013give} found that extending the choice set of the dictator by adding the possibility to take money from the recipient has the effect to make some dictators less likely to give. Therefore, these dictators preferred giving over keeping, when the taking option was not available, but preferred keeping over giving, when the taking option was available. This finding likewise cannot be explained by any preference over monetary payoffs. A conceptually similar point was also made by Krupka and Weber \cite{krupka2013identifying} and Capraro and Vanzo \cite{capraro2019power}, who found that even minor changes in the instructions of the dictator game can notably impact people's behaviour.

In the years after 2013, the inability of purely monetary-based models to explain empirically observed behaviour engulfed many other games and decision problems, whose experimental regularities had been previously thought to be explainable in terms of social preferences. Examples included the prisoner's dilemma \cite{kimbrough2016norms, capraro2018right}, the trust game \cite{kimbrough2016norms}, as well as different variants of the trade-off game \cite{capraro2018right, tappin2018doing, huang2019choosing}, thus resulting in a crisis of the social preference hypothesis.

\section{The rise of the moral preference hypothesis}\label{norm}

To solve a crisis, one needs a paradigm shift. The shift started in 2010, when Bicchieri and Chavez \cite{bicchieri2010behaving} proposed an elegant solution for one of the aforementioned empirical observations. This solution builds on classic work suggesting that, in everyday life, people's behaviour is partly determined by what they believe to be the norms in a given context \cite{smith2010theory,durkheim2017regles, parsons1937structure, geertz1973interpretation, schwartz1977normative, elster1989social, cialdini1990focus, bicchieri2005grammar,bicchieri2016norms,hawkins2019emergence}. This observation led behavioural scientists to propose several classifications of norms. Particularly relevant for the thesis of this review is the distinction between personal and social norms \cite{schwartz1977normative}. And moreover, among the social norms, the distinction between injunctive and descriptive norms \cite{cialdini1990focus}. Personal norms refer to internal standards about what is right or wrong in a given situation; injunctive norms refer to what other people approve or disapprove of in that situation; descriptive norms refer to what other people actually do. In one-shot anonymous games, like the games considered in this review, the distinction among personal, descriptive, and injunctive norms roughly corresponds to Bicchieri's personal normative beliefs, empirical expectations, and normative expectations \cite{bicchieri2005grammar}. See Table 3 for precise definitions.

The groundbreaking intuition of Bicchieri and Chavez \cite{bicchieri2010behaving} was to apply the theory of norms to deviations from monetary-based social preferences in the ultimatum game. Specifically, Bicchieri and Chavez showed that the ultimatum game offer that is consider to be fair by responders depends on the choice set available to the proposer; moreover, responders tend to reject offers that they consider unfair. This suggests that altruistic punishment is driven by responders following their personal norms, beyond the monetary consequences that these actions bring about. In particular, this explains the aforementioned results of Falk, Fehr, and Fischbacher \cite{falk2003nature}, that responders reject the same offer at different rates depending on the other offers available to the proposer.

Shortly after, in 2013, Krupka and Weber \cite{krupka2013identifying} applied a similar approach to several variants of the dictator game. However, instead of focusing on personal norms, they focused on injunctive norms. For each of the available actions, subjects were asked to declare whether they found the corresponding action to be ``very socially inappropriate'', ``somewhat socially inappropriate'', ``somewhat socially appropriate'', or ``very socially appropriate''. Subjects were given a monetary prize if they matched the modal choice made by other participants. Observe that, in this way, Krupka and Weber incentivised the elicitation of the injunctive norm. (The elicitation of personal norms cannot be incentivised.) In doing so, Krupka and Weber found that people believe that others think that avoiding a dictator game interaction is far less socially inappropriate than keeping the whole amount of money in a dictator game that one is obliged to play. Therefore, the empirical results summarised above regarding dictator games with an exit option \cite{dana2006you,lazear2012sorting} can be explained simply by a change in the perception of what is the injunctive norm in that context. Similarly, Krupka and Weber found that people believe that others think that keeping the money in a dictator game with a taking option is far less socially inappropriate than keeping the money in the dictator game without the taking option. In this way, they could explain also the results of List \cite{list2007interpretation}, Bardsley \cite{bardsley2008dictator}, and Cappelen et al. \cite{cappelen2013give} in terms of a change in the perception of the injunctive norm. Finally, Krupka and Weber presented a novel experiment in which subjects played the dictator game in either of two variants: in the Standard variant, dictators started with \$10 and had to decide how much of it, if any, to give to the recipient; in the Bully variant, the money was initially split equally among the dictator and the recipient, and the dictator could either give, take, or do nothing. The authors found that people were more altruistic in the Bully variant compared to the Standard variant, and this was driven by the fact that people rated ``taking from the recipient'' far less socially appropriate than ``not giving to the recipient''.

The work of Krupka and Weber suggests that taking into account injunctive norms is important to explain deviations from social preference models in the dictator game. But are the injunctive norms really the main force behind the observed behavioural changes, or are there also other norms playing more primary roles? In the last five years, a set of empirical studies tried to address this question. Schram and Charness \cite{schram2015inducing} analysed the behaviour of dictators who were given an advice from third parties about the injunctive norm. They observed that dictators became more pro-social only when their choices were made public. By contrast, when their choices remained private, they found no significant increase in pro-sociality, compared to the case in which they did not receive any information about the injunctive norm. These results indicate that, although injunctive norms might correlationally explain behavioural changes in anonymous (and thus private) dictator game experiments, they are unlikely to be the primary motivation. In fact, being that these games were played anonymously, in front of the screen of a computer, the intuition suggests that the norms primarily at play are the personal norms. Two recent works provide evidence for this hypothesis. Capraro and Vanzo \cite{capraro2019power} found that framing effects in the dictator game generated by morally loaded instructions can be explained by changes in the perception of what people ``personally think to be the right thing'' in the given context (i.e., their personal norms). Capraro et al. \cite{capraro2019increasing} showed that making personal norms salient prior to playing the dictator game (by asking subjects to state what they personally think to be the morally right thing to do) has a strong effect on subsequent dictator game donations, even persisting to a second-stage prisoner's dilemma interaction.

This set of works thus suggests that dictator game giving is driven by personal norms. Putting this together with the results of Bicchieri and Chavez, we obtain that both altruism and altruistic punishment can be explained by people following their personal norms.

More recently, this finding has been not only replicated, but, more importantly, also extended to explain several other forms of unselfish behaviour. In 2016, Kimbrough and Vostroknutov \cite{kimbrough2016norms} introduced a task ``that measures subjects' preferences for following rules and norms, in a context that has nothing to do with social interaction or distributional concerns''. They found that this measure of norm-sensitivity predicts dictator game altruism, trust game trustworthiness (but not trust), and ultimatum game rejection thresholds (but not offers). Taken together, this indicates that altruism, trustworthiness, and altruistic punishment are driven by a common desire to adhere to a personal norm. In 2017, Eriksson et al. \cite{eriksson2017costly} conducted an ultimatum game experiment under two different conditions. The difference, however, was only in the labels that were used to describe the action of refusing the proposer's offer. In one treatment, this action was labeled ``rejecting the proposer's offer'', while in the other treatment, the same action was labeled ``reducing the proposer's payoff''. Since these two options are monetarily equivalent, any utility function depending only on the monetary payoffs of the available actions predict that responders should behave the same way in both cases. But contrary to this prediction, Eriksson et al. found that responders displayed higher rejection thresholds in the ``rejection frame'' than in the ``reduction frame''. Moreover, they showed that the observed framing effect could be explained by a change in what people think to be the right thing to do. Specifically, subjects tended to rate the action of reducing the proposer's offer to be morally worse than the action of rejecting the proposer's offer, in spite of the fact that these two actions had the same monetary consequences. In 2018, Capraro and Rand \cite{capraro2018right} showed that behaviour in the trade-off game is highly sensitive to the labels used to describe the available actions. In line with Eriksson et al. \cite{eriksson2017costly}, Capraro and Rand also found that their framing effects could be explained by a change in what people think to be the right thing to do. Notably, framing effects in the trade-off game have been replicated several times \cite{tappin2018doing, huang2019choosing, capraro2019preferences,capraro2020gender,capraro2020does} and a recent work has shown that these moral framings tap into relatively internalised moral preferences \cite{capraro2020does}. Moreover, Capraro and Rand also considered a situation in which the personal norm conflicted with the descriptive norm, and found that people tend to follow the personal norm, rather than the descriptive norm. The same research also revealed a correlation between the framing effect in the trade-off game and giving in the dictator game and cooperation in the prisoner's dilemma, thus indicating that not only trade-off decisions are driven by personal norms, but that altruism and cooperation are also subject to that same facilitator. Cooperative behaviour is also typically correlated to altruistic behaviour \cite{capraro2014heuristics,peysakhovich2014humans,reigstad2017extending}, suggesting that they are driven by a common underlying motivation.

To the best of our knowledge, there are no works directly exploring the role of personal norms on truth-telling in the sender-receiver game. However, Biziou-van-Pol et al. \cite{biziou2015does} have shown that there is a positive correlation between truth-telling in the sender-receiver game (in the Pareto white lie condition), giving in the dictator game, and cooperation in the prisoner's dilemma, suggesting that these types of behaviours are driven by a common motivation. Since the aforementioned research suggests that altruism and cooperation are driven by personal norms, this correlation suggests that lying aversion is so too.

In sum, research accumulated in the last ten years suggests that several forms of one-shot, anonymous unselfishness, including altruism, altruistic punishment, truth-telling, cooperation, trustworthiness, and the equality-efficiency trade-off, can be explained using a unified theoretical framework, whereby people have moral preferences for following their personal norms, beyond the monetary payoff that these actions bring about. Of course, this is not meant to imply that monetary payoffs do not play any role in explaining one-shot unselfishness, but simply that something else, in addition to monetary payoffs, should be taken into account. The thesis is that this `something else' are the personal norms, which gives rise to the moral preference hypothesis as described in Table 4. Also, this is not meant to imply that other types of norms play no role in these forms of one-shot selfless behaviour. For example, nudging the injunctive norm in the prisoner's dilemma \cite{capraro2019increasing} and in the trade-off game \cite{human2020effect} has a similar effect as nudging the personal norm. Moreover, it is possible that social norms ultimately drive personal norms, because they allow to enhance or preserve one’s sense of self-worth and avoid self-concept distress, resulting in a self-reinforcing behaviour that eventually benefits one’s own self-image \cite{schwartz1977normative}. However, the aforementioned literature suggests that, at a proximate level, personal norms have a greater explanatory power, in the sense that they consistently explain people's behaviour also in games where injunctive norms have been shown to play a limited role (e.g., dictator game) or where descriptive norms play a limited role (e.g., the trade-off game).

\section{Practical applications}\label{applications}

Behavioural scientists and policy makers have been using norm-based interventions to foster pro-sociality in real life for decades \citep{bicchieri2009right, krupka2009focusing, zafar2011experimental, raihani2014dictator, d2017push, frey2004social, croson2010gendered, cialdini1991focus, ferraro2013using, agerstrom2016using, goldstein2008room, hallsworth2017behavioralist}. Although these paternalistic interventions have been criticised because they subtly violate people's freedom of choice \cite{hausman2010debate} and can be exploited by malicious institutions \cite{glaeser2005paternalism} (see \cite{sunstein2014nudge} for a response to these critiques), they are well-studied because, compared to standard procedures to foster pro-sociality (punishment and rewards), they allow to save the monitoring cost that the institution needs to pay in order to know who to punish or reward.

Norm-based interventions typically manipulate the descriptive or the injunctive norm in a given context, and show that this has an effect on people's behaviour in that same context. The more recent works reviewed in the previous section, showing that unselfish behaviour in one-shot, anonymous economic games is primarily driven by a desire to follow the personal norms, suggest that a more effective mechanism to increase pro-sociality might be to use norm-based interventions that target personal norms, rather than social norms. The interest in targeting personal norms, compared to other mechanisms to promote pro-sociality, is also that targeting personal norms is potentially cheaper than other mechanisms. Clearly, it is cheaper than punishment and rewards because it avoids the monitoring cost. Additionally, it saves the cost of collecting information about the behaviour or the moral judgments of other people, which forms the basis of interventions targeting social norms.

In recent years, there has been a growing body of research exploring the effect of nudging personal norms on various forms of unselfish behaviour. Some works using economic games found that making personal norms salient increases donations in the dictator game \cite{branas2007promoting,capraro2019increasing}, cooperation in the prisoner's dilemma \cite{dal2014right,capraro2019increasing}, as well as decreases in-group favouritism, at least on average \cite{bilancini2019right}. This suggests that nudging personal norms might be effective to increase pro-sociality in one-shot anonymous decisions that have consequences outside the laboratory. Along these lines, Capraro et al. \cite{capraro2019increasing} found that asking people to report what they personally think is the morally right thing to do increases crowdsourced charitable donations by 44\%.

\section{Models of moral preferences}\label{moral preferences}

We have thus seen that several forms of unselfish behaviour can be organised by moral preferences for following the personal norms. The question is, can we model this using a formal utility function?

There have been some attempts to formalise people's tendency to follow a norm \cite{benabou2006incentives, levitt2007laboratory, lopez2008aversion, andreoni2009social, dellavigna2012testing, kessler2012norms, alger2013homo, krupka2013identifying, kimbrough2016norms,kimbrough2020injunctive,kimbrough2020theory}. Most of these models, however, are either very specific in the sense that they can be applied only to certain games, or do not distinguish among different types of norms. Three models can be applied to every game of interest in this review (and, more generally, to every one-shot game) and distinguish among different types of norms.

Levitt and List \cite{levitt2007laboratory} introduced a model where the utility of an action $a$ depends on the monetary payoff associated to that action, $v_i(\pi_i(a))$, as well as on the moral cost (or benefit), $m(a)$, associated to that action:

$$u_i(a)=v_i(\pi_i(a))+m(a).$$

Levitt and List assumed that the moral cost (or benefit) depends primarily on three factors: whether the action is recorded or performed in the presence of an observer, whether the action has negative consequences on other players, and whether the action is in line with ``social norms or legal rules that govern behavior in a particular society''. Therefore, Levitt and List's model, although useful in many circumstances, does only mention social norms, while ignoring the effect of personal norms.

A similar model was considered by Krupka and Weber \cite{krupka2013identifying}, with the key difference that they focused on injunctive norms specifically. Krupka and Weber introduced a function $N$ defined over the set of available actions that, given an action $a$, returns a number $N(a)$ representing the extent to which society views $a$ as socially appropriate. They also assumed that people are heterogeneous in the extent to which they care about doing what society considers to be appropriate. In doing so, they obtain the utility function:

$$u_i(a)=v_i(\pi_i(a))+\gamma_i N(a).$$

As mentioned above, one of the main contributions of Krupka and Weber was to introduce an experimental technique to elicit the injunctive norm. To this end, they asked participants to rate each of the available actions in terms of their social appropriateness. Participants were incentivised to match the modal choice of the other participants.

Very recently, in 2020, Kimbrough and Vostroknutov presented a different approach, but still based on injunctive norms \cite{kimbrough2020theory}. Specifically, they introduced the utility function

$$u_i(a)=v_i(\pi_i(a))+\phi_i\eta(a),$$

where $\phi_i$ represents the extent to which $i$ cares about following the injunctive norm, and $\eta(a)$ represents a measure of whether the society thinks that $a$ is socially appropriate. Although this utility function looks very similar to the one proposed by Krupka and Weber, it differs from it in one important dimension. While Krupka and Weber's social appropriateness, $N(a)$, is computed by asking participants what they think others would approve or disapprove (and therefore it need not depend only on the monetary consequences of the available actions), Kimbrough and Vostroknutov's injunctive norm, $\eta$, is built axiomatically from the game and it is assumed to be inversely proportional to the overall dissatisfaction of the players, defined as the difference between what they get in a given scenario and what they could have gotten in others. This implies that one limitation of this approach is that people always prefer Pareto dominant allocations over Pareto dominated ones. But, in experiments, this property is not always satisfied. For example, when lying is Pareto dominant, some people still tell the truth, and these people tend to cooperate in a subsequent prisoner's dilemma and give in a subsequent dictator game \cite{biziou2015does}. Moreover, in trade-off games framed in such a way that the Pareto dominant allocation is presented as morally wrong, people tend to choose the Pareto dominated option \cite{capraro2018right,tappin2018doing}.

In sum, previous formal models consider only social norms or, more specifically, injunctive norms. But, as we have seen in the previous sections, unselfish behaviour in one-shot anonymous interactions is often driven by personal norms, rather than by social norms. Taking inspiration from the above models, one can formalise this using the utility function:

$$u_i(a)=v_i(\pi_i(a))+\mu_i P_i(a),$$

where $\mu_i$ represents the extent to which player $i$ cares about doing what s/he personally thinks to be the morally right thing to do and $P_i(a)$ represents the extent to which $i$ personally thinks that $a$ is morally right. This functional form might superficially seem similar to the ones discussed earlier, but it differs from those in two important points. One point is that the personal norm $P_i(a)$ typically depends on the individual $i$, whereas the injunctive norm depends on the society and the culture in which the individual is embedded. The second point is the very fact that $P_i$ represents the extent to which $i$ thinks that $a$ is the morally right thing to do, whereas $m(a), N(a)$, and $\eta(a)$ represent social norms. In general, the personal norm might not be aligned with the social norms. In practice, $P_i(a)$ can be estimated using a suitable experiment, whereas $\mu_i$ and $v_i$ can be estimated, on average, using statistical techniques, following a similar method as the one developed by Krupka and Weber for injunctive norms \cite{krupka2013identifying}. Specifically, one can estimate $P_i(a)$ by asking subjects to self-report the extent to which they personally think that action $a$ is the morally right thing to do. Then one can use these ratings to predict the behaviour, using a simple regression. The coefficient of this regression will give the average of the $\mu_i$'s. Also, putting the monetary payoffs in the regression, one can also get an estimation for the average of the $v_i$'s.

This utility function based on personal norms has a greater predictive power than its counterparts based only on social norms, in the sense that it explains behaviour in a larger set of games, compared to their counterparts based on social norms. We have seen earlier that Schram and Charness \cite{schram2015inducing} found that making the injunctive norm salient does \emph{not} increases altruistic behaviour in the anonymous dictator game. D'Adda et al. \cite{d2017push} found that making the descriptive norm salient has only a marginally significant effect on anonymous dictator game giving; this effect also vanishes in a second interaction, played immediately after. Along the same lines, Dimant, van Kleef and Shalvi \cite{dimant2019requiem} found that promoting the injunctive norm and promoting the descriptive norm does \emph{not} have any effect on people's honesty in a deception game in which subjects can lie for their benefit. On the other hand, numerous works have shown that nudging personal norms impacts several forms of unselfish behaviour, ranging from altruism \cite{branas2007promoting,capraro2019increasing}, altruistic punishment \cite{eriksson2017costly}, cooperation \cite{dal2014right,capraro2019increasing}, and the equality-efficiency trade-off \cite{capraro2018right}. Moreover, the effect typically persists for at least another interaction and even spills across contexts \cite{capraro2019increasing}. All these results are consistent with a utility function based on personal norms and are not consistent with a utility function based only on social norms.

We present a summary of all above-discussed moral preference models in Table 5.

\section{Future work}\label{future work}

This is an exciting field of research, which provides a unified view of human choices in several contexts of decision-making, while having, at the same time, significant practical implications. Nonetheless, there are several questions that need to be explored in future research, as detailed in what follows and summarised in Table 6.

\subsection{The utility function}

From a mechanistic perspective, the moral preference hypothesis raises the question of how can we express the utility function of a decision maker. Scholars have tried to give mathematical sense to people's morality since the foundation of mathematical economics \cite{jevons1879theory,bentham1996collected}. About two centuries later, the question is still open, even in the simple setting of one-shot anonymous interactions. One simple way to do so is to assume that people are torn between maximising their monetary payoff and doing what they personally think to be the morally right thing. This can be done with a utility function of the shape $u_i(a) = v_i(\pi_i(a))+\mu_i P_i(a)$. Although this utility function outperforms their counterparts based on social norms, as well as social preferences, it undoubtedly represents just a first candidate. Future work should explore other ways to formalise moral preferences, through finer experiments with the power to detect small variations in how people weight their personal norm against monetary incentives. Future work should also find ways to estimate what people think to be the right thing in a given context, without asking it to the participants in a separate experiment. The literature reviewed above shows that, in many cases, it is enough to change only one word in the instructions of a decision problem to change people's perception of what is the right thing to do in a given context. This suggests that $P_i(a)$ partly depends on the language in which the action $a$ is presented. Exploring this dependence can greatly improve the predictive power of the utility function. How can one do so? Recent work shows that emotional content in messages increases their diffusion in social media \cite{brady2017emotion, brady2019ideological, brady2019mad}. Translating this finding in the context of one-shot games, it suggests that the emotions carried by the instructions of the decision problem might contribute to the computation of $P_i$. Along these lines, it is possible that one can use sentiment analysis to better estimate $P_i$. Sentiment analysis is a technique developed by computational linguists that allows to assign a polarity to any given piece of text \cite{pang2002thumbs}. In principle, this polarity could enter the utility function of a decision maker and work as an additional motivation or obstacle for choosing an action, beyond its monetary consequences. In any case, mathematically describing or at least quantifying the seemingly intangible moral preferences, and in doing so building bridges between computational linguistics, behavioural economics, and moral psychology, is a fascinating direction for future work.

\subsection{Evolution of norms}

Where do personal norms come from? One explanation is that they come from the internalisation of behaviours that, although not individually optimal in the short term, are optimal in the long run. It is therefore important to understand which unselfish behaviours can be selected in the long term, and under which conditions. A promising line of research uses evolutionary game theory and statistical physics to find the conditions that promote the evolution of cooperation on networks \cite{perc_pr17}. More recently, scholars have started applying similar techniques also to study the evolution of other forms of unselfish behaviour \cite{capraro_fp18}, such as truth-telling in the sender-receiver game \cite{capraro2019evolution,capraro_pre20} and trustworthiness in the trust game \cite{kumar2020evolution}. Some works along this line have also looked at the evolution of choices in the ultimatum game \cite{page_prsb00, killingback_prsb01, iranzo_pone12, szolnoki_prl12}. Future work should extend the same techniques to other forms of unselfish behaviour.

\subsection{Personal norms versus social norms}

The experimental literature reviewed in the previous sections suggests that several forms of one-shot, anonymous unselfishness can be unified under a framework according to which people have preferences for following their personal norms. Moreover, preliminary evidence suggests that nudging personal norms can be an effective tool for fostering pro-sociality: making personal norms salient affects altruism, cooperation, altruistic punishment, and trade-off decisions between equality and efficiency \cite{branas2007promoting,capraro2019increasing,eriksson2017costly,dal2014right}.

This, of course, does not mean that the social norms play no role at all. For example, nudging injunctive norms has a significant effect on the one-shot, anonymous, prisoner's dilemma \cite{capraro2019increasing} and the trade-off game \cite{human2020effect}. One question that is still open, however, is whether these effects are fundamentally distinct from the effect of nudging personal norms. It is indeed possible that nudging injunctive norms in these games also nudge personal norms, and this is what makes people change their behaviour. A working paper suggests that people who follow injunctive norms in the trade-off game are different from those who follow personal norms \cite{human2020effect}. Therefore, it is possible that a larger model taking into account both personal and injunctive norms might have an even greater predictive power, at least in some contexts, than a model based exclusively on personal norms. Further experiments comparing the effect of nudging different norms are needed to clarify this point. The evidence in this case is indeed still lacunar. One study compared the relative effect of the descriptive and the injunctive norms in the dictator game, and found that people tend to follow the descriptive norm \cite{bicchieri2009right}. Another study compared the relative effect of nudging personal norms and the descriptive norms in the trade-off game, and found that people tend to follow the personal norms \cite{capraro2018right}. The aforementioned working paper compared the effect of nudging the personal and the injunctive norm in the trade-off game and found that they have a similar effect; moreover, when the two norms are in conflict, some people follow the personal norm and other follow the injunctive norm \cite{human2020effect}. This suggests that people's behaviour depends on their focus of attention within an interconnected matrix of norms. Therefore, future work should explore norm salience, also in cases where more than one type of norm is simultaneously made salient.

Research should also go beyond anonymous decisions and investigate what happens when choices are observable. The intuition suggests that when choices are observable, social norms may play a bigger role compared to when they remain private; in line with this intuition, Schram and Charness \cite{schram2015inducing} showed that nudging the injunctive norms impacts public but not private dictator game giving. However, no studies compared the relative effectiveness of targeting different norms in public decisions.

\subsection{Boundary conditions of interventions based on personal norms}

Having in mind potential practical applications, another important question concerns the boundary conditions of interventions based on personal norms. From a temporal perspective, previous research suggests that interventions targeting personal norms can last for several interactions within the same experiment \cite{dal2014right,capraro2019increasing}. However, it seems unrealistic to expect that their effect will last indefinitely. For example, a recent field experiment targeting injunctive norms found an effect that diminishes after repeated interventions, although it can be restored after waiting a sufficient amount of time between interventions \cite{ito2018moral}. From the decisional context point of view, there will certainly be behavioural domains in which targeting personal norms might not be as effective. For example, a recent work suggests that risky cooperation in the stag-hunt game is primarily driven by preferences for efficiency, rather than by preferences for following personal norms \cite{capraro2019preferences}.

\subsection{External validity of interventions based on personal norms}

Given the potential relevance of this line of work for the society at large, future studies should explore the external validity of interventions based on personal norms. At the time of this writing, only one study investigated the effect of nudging personal norms in contexts in which decisions have consequences outside the laboratory. This study found that nudging personal norms increases crowdsourced charitable donations to real humanitarian organisations by 44\% \cite{capraro2019increasing}.

\subsection{The moral phenotype and its topology}\label{suse: phenotype}

We have seen that different forms of unselfish behaviour can be explained by a general tendency to do the right thing. We are tempted to call this tendency ``moral phenotype'', extending the notion of ``cooperative phenotype'' introduced by Peysakhovich, Nowak, and Rand \cite{peysakhovich2014humans}. See also \cite{reigstad2017extending}. In their work, Peysakhovich and colleagues observed that pro-social behaviours in the dictator game, the public goods game (a variant of the prisoner's dilemma with more than two players), and the trust game (both players) were all correlated; and they termed this general pro-social tendency cooperative phenotype. Therefore, the cooperative phenotype is uni-dimensional. On the other hand, the moral phenotype is likely to be multi-dimensional. For example, we have seen earlier that both altruistic punishment and altruistic giving are driven by preferences for doing the right thing. However, Peysakhovich, Nowak, and Rand \cite{peysakhovich2014humans} found that they are not correlated. It is possible that they are not correlated because they come from different personal norms. The multi-dimensionality of morality is not a new idea, and several authors have come to suggest it in the last decades from different routes. For example, Haidt and colleagues argue that differences in people's moral concerns can be explained by individual differences across six ``foundations'' \cite{haidt2004intuitive,graham2009liberals,haidt2012righteous}. Kahane, Everett and colleagues have shown that (act) utilitarianism decomposes itself in at least two dimensions \cite{kahane2018beyond,everett2020switching}. Curry, Mullins, and Whitehouse \cite{curry2019good} have reported that seven moral rules are universal across societies, but societies vary on how they rank them. However, we are not aware of any work exploring how different personal norms link to different forms of one-shot unselfishness.

Another topological property of the moral phenotype that deserves further scrutiny is the boundary. Does, for example, the moral phenotype include decisions that are strategically unselfish, such as strategic fairness (ultimatum game offers) and trust (trust game transfers), both of which maximise the decision maker's payoff depending on the decision maker's beliefs about the behaviour of the other player? Previous evidence is limited and mixed. Bicchieri and Chavez \cite{bicchieri2010behaving} showed that ultimatum game offers are partly driven by normative beliefs; Peysakhovich, Nowak, and Rand \cite{peysakhovich2014humans} found that trustees' decisions correlate with dictator game and public goods game decisions. By contrast, Kimbrough and Vostroknutov \cite{kimbrough2016norms} found that trustees' and proposers' decisions are not correlated to their measure of norm-sensitivity.

\subsection{A dual-process approach to personal norms}

Do personal norms come out automatically, or do they require deliberation? Research recently explored the cognitive basis of unselfish behaviour, by using cognitive process manipulation, such as time pressure and cognitive load, in order to favour instinctive responses \cite{rand_n12, andersen2018allowing, bereby2018honesty, bouwmeester2017registered, capraro2019time, capraro2017deliberation, chen2019cognitive, chuan2018field, everett2017deliberation, holbein2019insufficient}. It has been shown that promoting intuition favours cooperation \cite{rand2016cooperation} and altruistic punishment \cite{hallsson2018fairness}. The evidence regarding altruism is instead more mixed \cite{rand2016social,fromell2020altruism}. Instead, a meta-analysis suggests that intuition decreases truth-telling, when lying harms abstract others, while leaving it unaffected when it harms concrete others \cite{kobis2019intuitive}. Furthermore, results are inconclusive in the context of trustworthiness and the equality-efficiency trade-off (see \cite{capraro2019dual} for a review). This line of work suggests that whether personal norms come out automatically or require deliberation may not have a general answer, but might depend on the specific behavioural context, and possibly also on the individual characteristics of the decision maker. More work is needed to understand which personal norms, in which context, and for which people, become internalised as automatic reactions.

\section{Conclusions}\label{conclusion}

The moral preference hypothesis is emerging as a unified framework to explain a wide range of one-shot, anonymous unselfish behaviours, including cooperation, altruism, altruistic punishment, truth-telling, trustworthiness, and the equality-efficiency trade-off. This framework has promising practical implications, given that interventions making personal norms salient have been shown to be effective at increasing charitable donations. Future work should explore further mathematical formalisations of moral preferences in terms of a utility function, investigate the evolution and internalisation of personal norms, study the external validity and the boundary conditions of policy interventions based on personal norms, compare the relative effectiveness of targeting different types of norms, examine the topology of the moral phenotype, and analyse the cognitive foundations of morality, possibly using a dual-process perspective.

Overall, the goal of this line of research should be to build bridges between different scientific disciplines to arrive at a better, perhaps more mechanistic, explanation of human decision-making. The outlined mathematical formalism for morality should be used to inform future models aimed at better understanding selfless actions, and it should also be used in artificial intelligence to better navigate the complex landscape of human morality and to better emulate human decision-making. Ultimately, the goal is to use the obtained insights to develop more efficient policies and interventions to increase good virtues and decrease bad ones, and to collectively strive towards better human societies.

The past century has seen strict compartmentalisation of different scientific disciplines leading to groundbreaking and important discoveries that might had been impossible without it. But while technology and industry might fare well on idiosyncratic breakthroughs, human societies do not. The grandest challenges of today remind us that sustainable social welfare and organisation require a wholesome interdisciplinary and cross-disciplinary approach, and we hope this review will be an inspiration towards this goal.

\begin{acknowledgments}
This work was supported by the Slovenian Research Agency (Grant Nos. P1-0403, J1-2457, J4-9302, and J1-9112).
\end{acknowledgments}

\clearpage

\noindent\fcolorbox{black}{boxcolor}{%
\parbox{16cm}{%
\underline{\textbf{Table 1: Glossary of games and unselfish behaviours}}

\textbf{Dictator game}: We measure altruistic behaviour using the dictator game. The \emph{dictator} is given a certain amount of money and has to decide how much of it, if any, to give to the \emph{recipient}, who starts with nothing. The recipient is passive.

\textbf{Prisoner's dilemma}: We measure cooperative behaviour using the prisoner's dilemma. Two players simultaneously decide whether to cooperate or to defect. Cooperating means paying a cost $c$ to give a benefit $b>c$ to the other player; defecting means doing nothing.

\textbf{Sender-Receiver game}: We measure lying aversion using the sender-receiver game. The \emph{sender} is given a private information and has to report it to the \emph{receiver}. In some experiments the receiver is passive \cite{gneezy2013measuring,biziou2015does}, in others is active \cite{gneezy2005deception,erat2012white}. Here we focus on the case in which the receiver is passive. In this case, if the sender reports the truthful information, then the sender and the receiver are paid according to Option A; if the sender reports an untruthful information, then the sender and the receiver are paid according to Option B. Only the sender knows the exact payoffs associated to the two options. Depending on these payoffs, one can classify lies into four main classes: black lies are those that benefit the sender at a cost to the receiver; altruistic white lies are those that benefit the receiver at a cost to the sender; Pareto white lies are those that benefit both the sender and the receiver; spiteful lies are those that harm both the sender and the receiver.

\textbf{Trade-Off game}: We measure the trade-off between equality and efficiency using the trade-off game. A decision-maker has to decide between two possible allocations of money that affect people other than the decision-maker. One decision is equal (i.e., all people involved in the interaction receive the same monetary payoff), the other decision is efficient (i.e., the sum of the monetary payoffs of all people is greater than it is in the equal allocation).

\textbf{Trust game}: We measure trustworthiness using the second player in the trust game. The \emph{truster} is given a certain amount of money and has to decide how much of it, if any, to transfer to the \emph{trustee}. The amount sent to the trustee is multiplied by a constant (usually equal to 3) and given to the trustee. The trustee decides how much of the amount s/he received to return to the truster.

\textbf{Ultimatum game}: We measure altruistic punishment using the second player in the ultimatum game. The \emph{proposer} makes an offer about how to split a sum of money between him/herself and the \emph{responder}. The responder decides whether to accept or reject the offer. If the offer is accepted, the proposer and the responder get paid according to the agreed offer; if the offer is rejected neither the proposer nor the responder get any money. Rejecting a low offer is considered to be a measure of altruistic punishment.
}
}

\clearpage

\noindent\fcolorbox{black}{boxcolor}{%
\parbox{16cm}{%
\underline{\textbf{Table 2: Social preference models}}

Let $x_i$ be the monetary payoff of player $i$. Social preference models assume that the utility function of player $i$, $u_i$, is defined over the monetary payoffs that are associated with the available actions. The main functional forms that have been proposed are the following.

\textbf{Ledyard (1994)}: $u_i(x_1,\ldots,x_n)=x_i+\alpha_i\sum_{j\neq i}x_j$, where $\alpha_i$ is an individual parameter representing $i$'s level of altruism. People with $\alpha_i=0$ maximise their monetary payoff; people with $\alpha_i>0$ are altruistic; people with $\alpha_i<0$ are spiteful.

\textbf{Levine (1998)}: $u_i(x_1,\ldots,x_n)=x_i+\sum_{j\neq i}\frac{\alpha_i+\lambda\alpha_j}{1+\lambda}x_j$, where $\alpha_i$ is an individual parameter representing $i$'s level of altruism, whereas $\lambda\in[0,1]$ is a parameter representing how sensitive players are to the level of altruism of the other players.

\textbf{Fehr and Schmidt (1999)}: $u_i(x_1,\ldots,x_n)=x_i-\frac{\alpha_i}{n-1}\sum_{j\neq i}\max(x_j-x_i,0)-\frac{\beta_i}{n-1}\sum_{j\neq i}\max(x_i-x_j,0)$, where $\alpha_i,\beta_i$ are individual parameters representing the extent to which player $i$ cares about disadvantageous and advantageous inequities, respectively

\textbf{Bolton and Ockenfels (2000)}: $u_i(x_1,x_2)=\alpha_ix_i-\frac{\beta_i}{2}\left(\sigma_i-\frac12\right)^2$, where $\sigma_i=\frac{x_i}{x_1+x_2}$, with $\sigma_i=\frac12$ if $x_1+x_2=0$, $\alpha_i>0$ is an individual parameter representing the extent to which player $i$ cares about their own monetary payoff, and $\beta_i>0$ is an individual parameter representing the extent to which player $i$ cares about minimising the distance between their share and the fair share.

\textbf{Andreoni and Miller (2002)}: $u_1(x_1,x_2)=\left(\alpha_1 x_1^{\rho_1}+(1-\alpha_1)x_2^{\rho_1}\right)^{1/\rho_1}$, where $\alpha_1$ represents the extent to which the dictator cares about their own payoff, whereas $\rho_1$ takes into account a potential convexity in the preferences.

\textbf{Charness and Rabin (2002)}: $u_2(x_1,x_2)=(\rho_2 r+\sigma_2 s)x_1+(1-\rho_2 r - \sigma_2 s)x_2$. Depending on the relative relationship between $\rho_2$ and $\sigma_2$, this utility function can cover several cases, including competitive preferences, inequity aversion preferences, and social efficiency preferences.
}
}

\clearpage

\noindent\fcolorbox{black}{boxcolor}{%
\parbox{16cm}{%
\underline{\textbf{Table 3: The classification of norms}}

Behavioural scientists have long been aware of the fact that people's behaviour in a given context is influenced by what are perceived to be the norms in that context. In the same context, multiple norms might be at play. Scholars have proposed several norm classifications. In this review, we will be mainly concerned with the following three.

\textbf{Schwartz} \cite{schwartz1977normative} classified norms into two main categories, namely \emph{personal norms} and \emph{social norms}. Personal norms refer to internal standards about what is right and what is wrong in a given context. Social norms refer to rules and standards of behaviour that affect the choices of individuals without the force of law. Social norms are typically externally motivated.

\textbf{Cialdini, Reno and Kallgren} \cite{cialdini1990focus} focused on social norms and classified them into two main categories, namely \emph{injunctive norms} and \emph{descriptive norms}. Injunctive norms refer to what people think others would approve or disapprove. Descriptive norms refer to what others actually do.

\textbf{Bicchieri} \cite{bicchieri2005grammar} proposed a classification in three main categories, namely \emph{personal normative beliefs}, \emph{empirical expectations}, and \emph{normative expectations}. Personal normative beliefs refer to personal beliefs about what should happen in a given situation. Empirical expectations refer to personal beliefs about how others would behave in a given situation. Normative expectations refer to personal beliefs about what others think one should do.

Therefore, to the extent to which people believe that what should (or should not) happen in a given situation corresponds to their internal standards about what is right (or wrong), then Bicchieri's personal normative beliefs correspond to Schwartz's personal norms. In one-shot anonymous games (where decision makers receive no information about the behaviour of other people playing in the same role), descriptive norms correspond to empirical expectations (we replace the actual behaviour of others with the beliefs). Finally, normative expectations correspond to injunctive norms. Therefore, at least for the games and decision problems considered in this review, Bicchieri's classification can be interpreted as a synthesis of the previous two classifications.
}
}

\clearpage

\noindent\fcolorbox{black}{boxcolor}{%
\parbox{16cm}{%
\underline{\textbf{Table 4: The moral preference hypothesis}}

Previous work explained unselfish behaviour in one-shot, anonymous economic games using social preferences defined over monetary outcomes. According to this ``social preference hypothesis'', some people act unselfishly because they do not only care about their own monetary payoff, but they also care about the monetary payoffs of other people. However, especially in the last five years, numerous experiments challenged social preference models. The best way to organise these results is through the moral preference hypothesis, according to which people have preferences for following their own personal norms -- what they think to be the right thing to do -- beyond the monetary consequences that these actions bring about. This framework outperforms the social preference hypothesis at organising cooperation in the prisoner's dilemma, altruism in the dictator game, altruistic punishment in the ultimatum game, trustworthiness in the trust game, truth-telling in the sender-receiver game, and trade-off decisions between equality and efficiency in the trade-off game.
}
}

\clearpage

\noindent\fcolorbox{black}{boxcolor}{%
\parbox{16cm}{%
\underline{\textbf{Table 5: Moral preference models}}

Let $a$ be an action for player $i$. Moral preference models assume that the utility function of player $i$, $u_i$, describes a tension between the material payoff associated to $a$, $v_i(\pi_i(a))$, and the moral utility. The main functional forms that have been proposed are the following.

\textbf{Levitt and List (2007)}: $u_i(a)=v_i(\pi_i(a))+m(a)$. The moral cost or benefit associated to $a$, $m(a)$, is assumed to depend on whether the action is observable, on the material consequences of that action, and on the set of \emph{social norms} and rules in place in the society where the decision maker lives.

\textbf{Krupka and Weber (2013)}: $u_i(a)=v_i(\pi_i(a))+\gamma_i N(a)$, where $\gamma_i$ is the extent to which $i$ cares about following the \emph{injunctive norm} and $N(a)$ represents the extent to which society views $a$ as socially appropriate.

\textbf{Kimbrough and Vostroknutov (2020)}: $u_i(a)=v_i(\pi_i(a))+\phi_i\eta(a)$, where $\phi_i$ is the extent to which $i$ cares about following the \emph{injunctive norm} and $\eta(a)$ represents the extent to which society views $a$ as socially appropriate. (The main difference between $\eta(a)$ and $N(a)$ regards the way they are computed.)

\textbf{Our proposal:} $u_i(a)=v_i(\pi_i(a))+\mu_i P_i(a)$, where $\mu_i$ represents the extent to which $i$ cares about following their own \emph{personal norms} and $P_i(a)$ represents the extent to which $i$ personally thinks that $a$ is the right thing to do.

}
}

\clearpage

\noindent\fcolorbox{black}{boxcolor}{%
\parbox{16cm}{%
\underline{\textbf{Table 6: Outstanding challenges}}

\begin{itemize}
\item Exploring in which contexts interventions targeting personal norms are more effective at promoting one-shot unselfish behaviour than interventions targeting social norms.
\item Finding the boundary conditions of interventions targeting personal norms.
\item Investigating the dimension and the boundary of the ``moral phenotype'', to understand how different personal norms can drive different forms of unselfish behaviour and whether the moral phenotype includes behaviours that are strategically unselfish, such as strategic fairness and trust.
\item Building bridges between computational linguistics, moral psychology, and behavioural economics, with the goal of understanding how to express people's utility function also in terms of the instructions of a decision problem.
\item Using techniques from evolutionary game theory, applied mathematics, network science, and statistical physics to explore which types of unselfish behaviour are more likely to evolve in order to understand which personal norms are more likely to be internalised.
\item Exploring the cognitive basis of personal norms using a dual-process perspective.
\end{itemize}
}
}

\end{document}